\documentclass[a4paper,12pt]{article}
\usepackage{mathpple}   
\usepackage[T1]{fontenc}
\usepackage[ansinew]{inputenc}
\usepackage{fancyhdr}
\usepackage{amstext} \usepackage{amssymb}
\usepackage{epsfig}
\usepackage{graphicx}
\usepackage{latexsym}
\usepackage{dcolumn}
\usepackage{citesort}
\usepackage{bm}
\usepackage{rotating}

\newcommand{\pbar}{\mbox{$\overline{\mathrm p}$}}
\newcommand{\pbhe}{$\overline{\rm p} {\rm He}^{+}$}
\newcommand{\pbheion}{$\overline{\rm p} {\rm He}^{2+}$}

\newcommand{\Se}{$\vec{S}_e$}
\newcommand{\Lp}{$\vec{L}_{\overline{p}}$}
\newcommand{\Sp}{$\vec{S}_{\overline{p}}$}

\newcommand{\mup}{$\vec{\mu}_{\overline{p}}$}

\newcommand{\gl}{$g^{\overline{p}}_{\ell}$ }
\newcommand{\gs}{$g^{\overline{p}}_{\mathrm{s}}$ }
\newcommand{\nuHF}{$\nu_{\mathrm{HF}}$}
\newcommand{\DnuHF}{$\Delta \nu_{\mathrm{HF}}$}
\newcommand{\nuHFp}{$\nu_{\mathrm{HF}}^+$}
\newcommand{\nuHFm}{$\nu_{\mathrm{HF}}^-$}
\newcommand{\nuSHF}{$\nu_{\mathrm{SHF}}$}
\newcommand{\nuSHFp}{$\nu_{\mathrm{SHF}}^+$}
\newcommand{\nuSHFm}{$\nu_{\mathrm{SHF}}^-$}

\begin{document}

\title{Precision spectroscopy of 
antiprotonic helium}
\author{E. Widmann\\
 {\em Stefan Meyer Institute for Subatomic Physics}\\
 {\em Boltzmanngasse 3, 1090 Vienna, Austria}}

\date{ }
\maketitle

\begin{abstract}
Antiprotonic helium, a neutral exotic three-body system consisting of a helium nucleus, an electron and an antiproton, is being studied at the Antiproton Decelerator of CERN by the ASAUCSA collaboration. Using laser spectroscopy of the energy levels of the antiproton in this system and comparison to theory, a value of the antiproton-to-electron mass ratio with an error of 3 ppb could be obtained. This result agrees with the most precise measurement of the value of the proton and allows us to extract a limit of the equality of the proton and antiproton charge and mass of 2 ppb. Using microwave spectroscopy, the hyperfine structure of antiprotonic helium has been measured to 30 ppm. Experimental  improvements are expected to soon provide a new value for the magnetic moment of the antiproton.
\end{abstract}


\section{Introduction}

\begin{figure}
\centering
\includegraphics[width=0.9\textwidth]{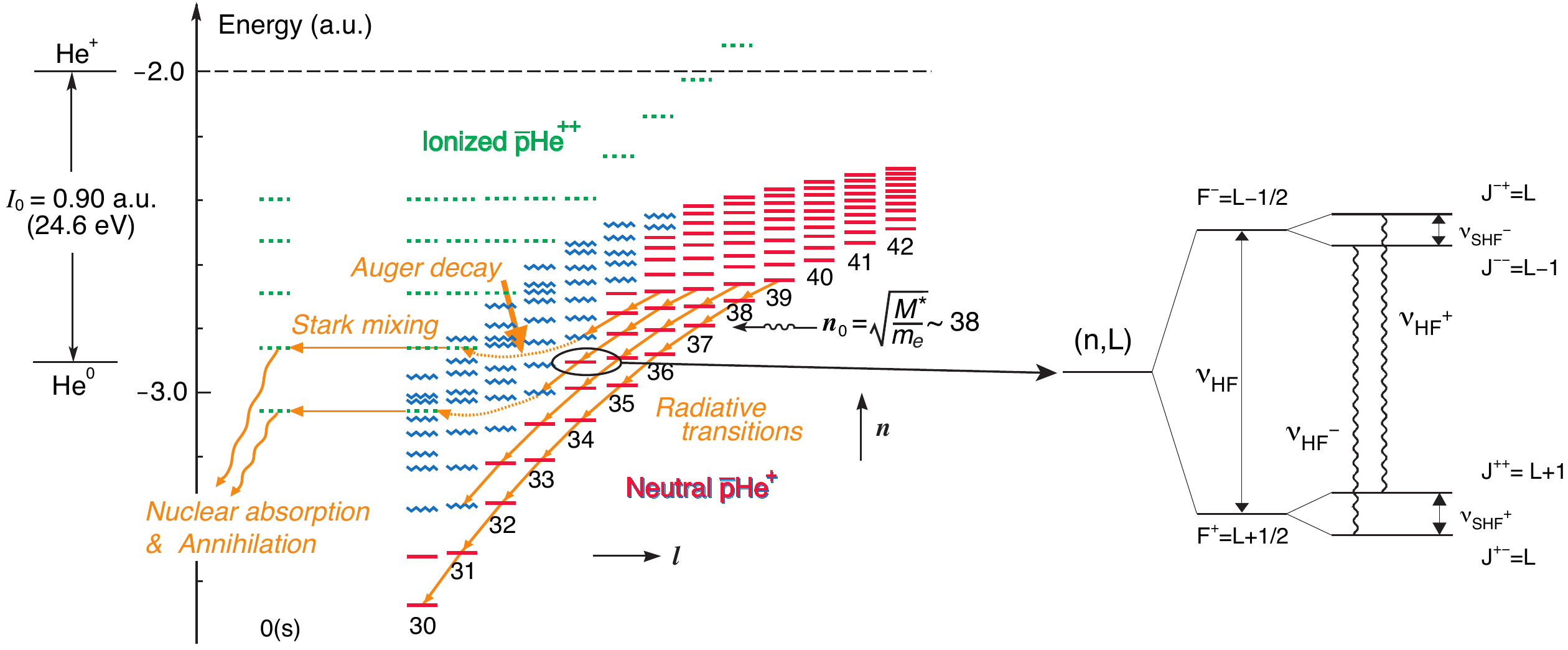}
\caption{Left: Coulomb energy levels of antiprotonic helium. The energy diagram is divided into metastable levels (red) which decay via radiative transitions, and short-lived ones (blue) where Auger decay dominates. Subsequently the \pbar\ is transferred to low-$l$ states by Stark mixing, from where it annihilates with one of the nucleons. Right: Hyperfine splitting of a Coulomb-level $(n,L)$, with $L$ being the antiproton angular momentum. }
\label{EWfig:leveldiag}       
\end{figure}

Initially discovered at KEK in 1991 \cite{Iwasaki:91}, antiprotonic helium has been studied extensively at LEAR \cite{Yamazaki:01} and the Antiproton Decelerator (AD) of CERN. It is an exotic 3-body system \pbar\  -- $e^-$ -- He$^{2+}$ or short \pbhe, where the antiproton occupies highly excited metastable states with principal quantum number $n$ and angular momentum quantum number $l$ of $(n,l)\sim 33\ldots39$. Its average lifetime of $\sim 3$ $\mu$s allows access to the antiproton energy levels by laser and microwave spectroscopy. The energy levels of \pbhe\ are shown in Fig.~\ref{EWfig:leveldiag}. Using laser spectroscopy, the levels $(n,l)$ have been determined with high precision which allows a determination of the antiproton charge and mass as described below. Microwave spectroscopy gives access to the hyperfine structure and therefore to the magnetic moment of the antiproton, a quantity that is known so far to only 0.3\%.

\begin{figure}[b] 
   \centering
   \includegraphics[width=9cm]{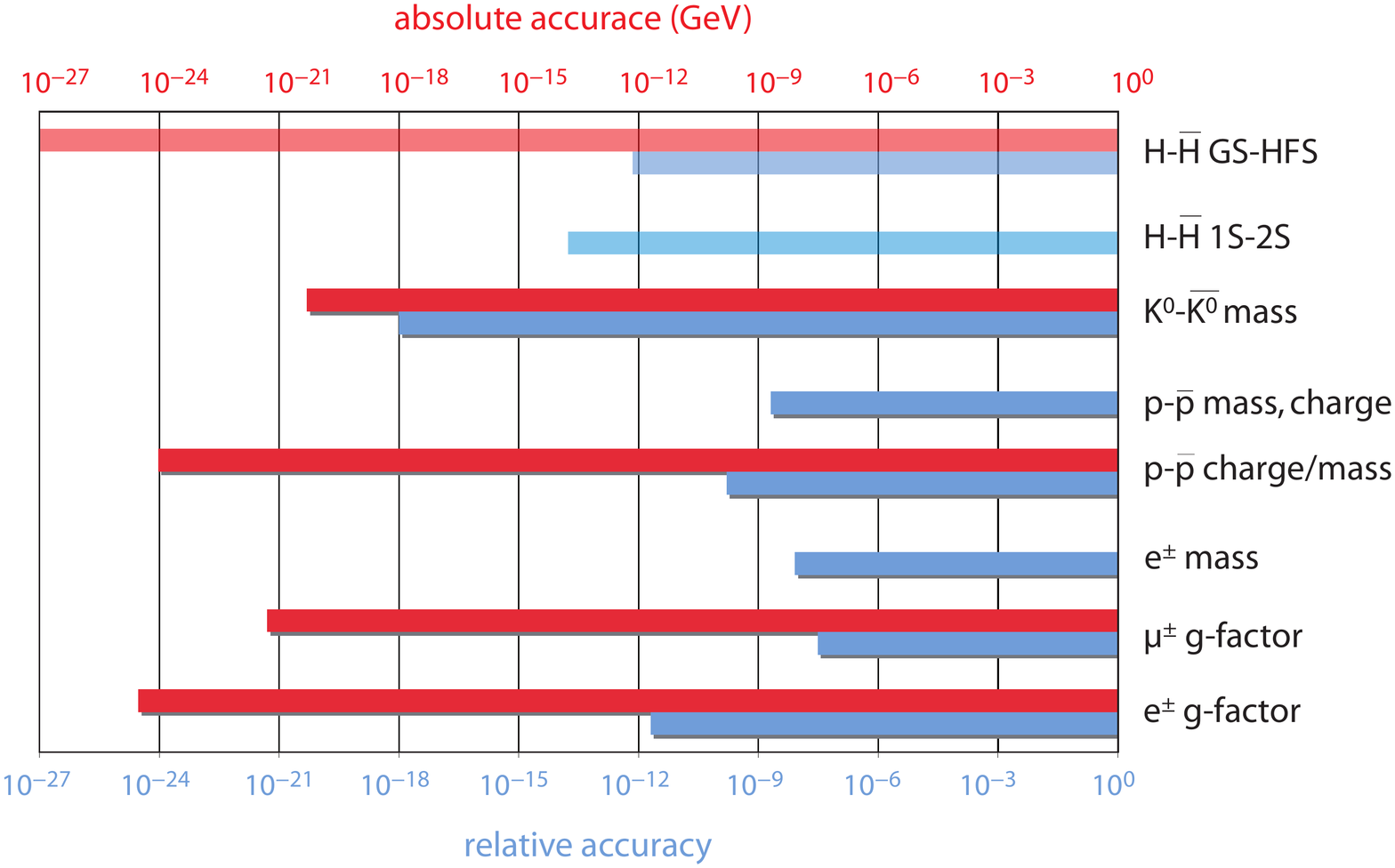} 
   \caption{Comparison of different CPT-tests in terms of absolute and relative precision. For details see text.}
   \label{EWfig:CPT}
\end{figure}

Both measurements constitute a precise test of CPT symmetry in the baryon sector. Fig.~\ref{EWfig:CPT} compares such tests for different cases including predictions for antihydrogen (light colours) which will not be discussed here. The blue bars correspond to values which can be extracted from the particle data group compilation \cite{PDG:06}: shown are the {\em relative} differences between particle and antiparticle properties. This way of comparing different tests of CPT, however, is misleading as quite different quantities are compared for different sectors (leptons, mesons, baryons, atoms!) and it is known that {\em e.g.} CP violation  only occurs in the meson sector. A better way of comparing CPT tests in different sectors has first been suggested in the framework of the Standard Model Extension (SME) by Kostelecky et al. \cite{Colladay:97}. In his phenomenological model, CPT and Lorentz invariance violating terms are added to the Dirac equation to describe possible symmetry breakings. This implies that the parameters, which can be determined from experiment, have an {\em absolute} dimension of energy. Kostelecky and his groups analyzed many possible experiments and arrived at the limits which are shown as red bars in Fig.~\ref{EWfig:CPT} \cite{Bluhm:1997lr,Bluhm:1999qy,Kostelecky:1998uq,Bluhm:2000fj}.  In this context it becomes evident that the most precise tests of CPT are measurements involving atomic physics techniques, where small quantities on an energy scale are measured to extremely high precision.

\section{Charge and mass of the antiproton}

The fact that the \pbar\ survives for several microseconds in \pbhe\ while undergoing radiative transitions whose wavelengths lie in the visible region made it well suited for laser spectroscopy. The technique developed called ``forced annihilation'' made use of the additional fact that antiprotons mostly follow cascades with $\Delta n = \Delta l = -1$ or $\Delta v = 0$ with $v=n-l-1$ being the vibrational quantum number. At the end of each such cascade of Fig.~\ref{EWfig:leveldiag} there is a pair of adjacent metastable and short-lived states. When the laser is on resonance, the antiproton is transferred to the short-lived state and annihilates immediately, leading to a sharp spike in the DATS (Delayed Annihilation Time Spectrum). 

The experimental accuracy has been steadily improved, from 0.5 ppm achieved at the end of LEAR to currently 2 ppb. A first major improvement came from the construction of a radio frequency quadrupole decelerator, RFQD, in a joint venture of the ASACUSA collaboration and the CERN PS division \cite{RFQD-01}. This RFQD decelerates the AD beam from 5 MeV to 60--120 keV, thus allowing to stop antiprotons in helium gas of three orders of magnitude less density. This way, the dominating systematic error of density shifts of the transition wavelengths could be avoided. By comparing the experimental results with two independent theoretical calculations, a CPT limit of 10 ppb on the maximum relative difference of charge and mass of the antiproton could be established \cite{Hori:03} as discussed below.

\begin{figure}
\centering
\includegraphics[width=6.5cm]{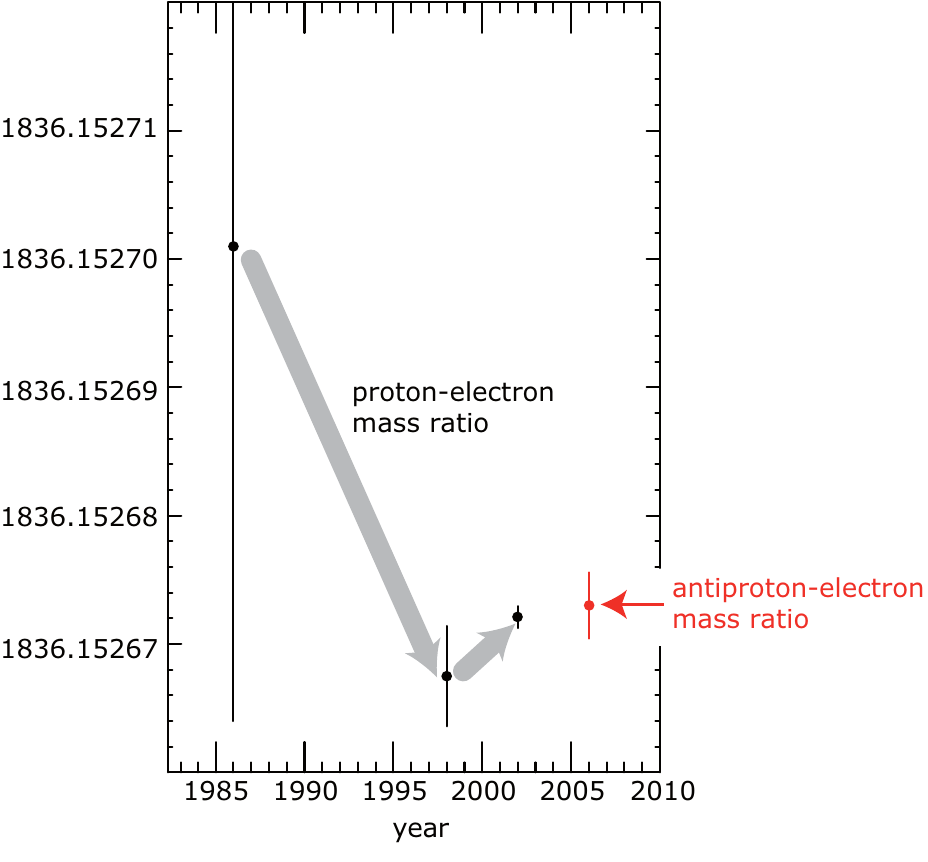}
\includegraphics[width=6.5cm]{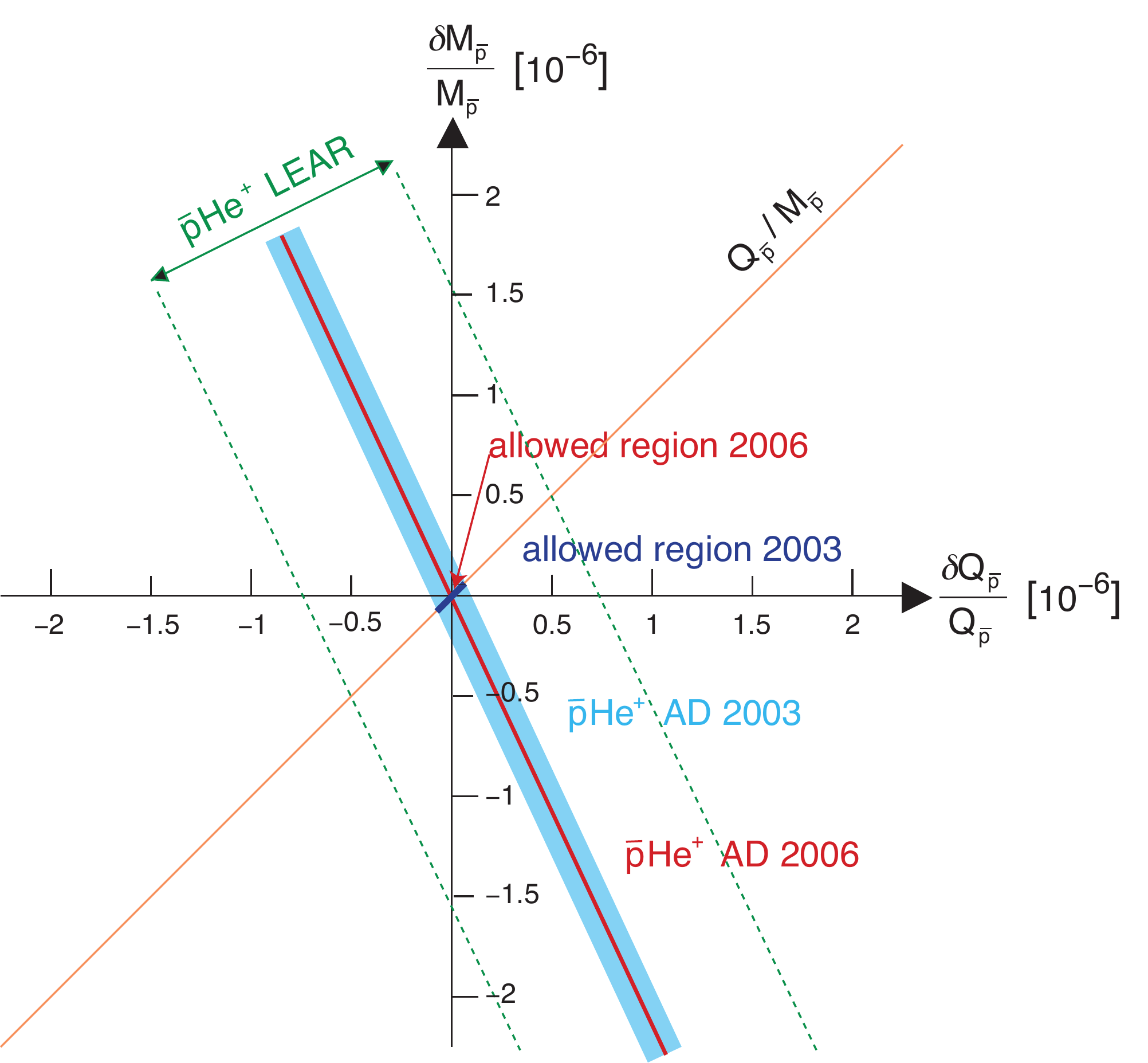}
\caption{Left: proton and antiproton-to-electron mass ratio. Right: Combining two measurements of TRAP ($Q/M$) and ASACUSA ($MQ^2$) to extract individual limits on antiproton and proton charge and mass.}
\label{EWfig:m-vs-exp}       
\end{figure}

The next step in accuracy came with utilizing pulse-amplified cw lasers that were locked to a frequency comb. With this method, the two now dominant sources of systematic errors: the laser band width and the absolute calibration of its wavelength, could be strongly reduced. Again the same 13 transitions were measured and compared to the remaining theory that agrees well with the data. By taking the antiproton-to-electron mass ratio $M_{\overline{\mathrm p}} / m_{\mathrm e}$ as a free parameter in the theory and fitting all data points for the various transitions, a value of
$M_{\overline{\mathrm p}} / m_{\mathrm e} = 1836.152\,674\, (5)$
was deduced, which is in good agreement with the latest proton value from CODATA (cf. Fig.~\ref{EWfig:m-vs-exp} left) \cite{Hori:06}. The error on this figure is about 3 ppb.

This treatment assumes, that the charge of proton and antiproton have the same value of $|e|$. However, as was pointed out by Hughes and Deutch earlier \cite{Hughes:92}, the independent limits on the charge of p and \pbar\ are much less, about $|Q_{\mathrm p} - Q_{\overline{\mathrm p}}|/e < 2\times 10^{-5}$. In order to extract independent limits for both charge and mass, one can go the following way: The TRAP collaboration at LEAR \cite{Gabrielse:99} has measured $Q/M$ to  a precision of $\sim 10^{-10}$. The measured transitions wave length of \pbhe\ are dominated by the Rydberg constant which is $\propto MQ^2$. As layed out in \cite{Torii:99,Yamazaki:01}, one can use both measurements to get separate limits of $Q$ and $M$.
Fig.~\ref{EWfig:m-vs-exp} right illustrates this procedure: the allowed region for charge and mass of the antiproton is where the two areas of opposite-sign slope intersect. 

In general, each measured transition frequency depends in a slightly different way on charge and mass than in the simple model used above. The relation becomes

\begin{equation}
\delta_{\mathrm p}= \frac{Q_{\overline{\mathrm p}}+Q_{\mathrm p}}{Q_{\mathrm p}}
\sim \frac{M_{\mathrm p}-M_{\overline{\mathrm p}}}{M_{\mathrm p}}
= \frac{1}{f} \frac{\nu_{\mathrm th}-\nu_{\mathrm exp}}{\nu_{\mathrm exp}},
\end{equation}

with $f=2-5$ being a transition-dependent ``sensitivity factor'' obtained by theory. The final result is obtained by averaging over all measured transitions. 
Since the TRAP measurement of $Q/M$ is still one order of magnitude more precise the the \pbhe\ spectroscopy, the errors on the individual properties is given by the ASACUSA error of 2 ppb \cite{Hori:06}.

A further improvement from this point is only possible by using two-phtoton spectroscopy, since the line width is now dominated by Doppler broadening. A first test was done in 2006 and the results look encouraging.

\section{Magnetic moment of the antiproton}

The energy levels in antiprotonic helium which were so far denoted by principle $n$ and angular momentum quantum number $l$ exhibit a hyperfine (HF) structure due to the magnetic interaction of the constituents of \pbhe. Since the electron in \pbhe\ is predominantly in the ground state, the total angular momentum $\vec{l}$ is equal to the one of the antiproton, \Lp. The magnetic moment of \pbar\ is given by the sum of angular and spin terms
\begin{equation}
\vec{\mu_{\overline{\mathrm p}}}=(g^{\overline{\mathrm p}}_\ell \vec{L_{\overline{\mathrm p}}} + g^{\overline{\mathrm p}}_s \vec{S_{\overline{\mathrm p}}}) \mu_{\overline{\mathrm{N}}},
\end{equation}
where $\mu_{\overline{\mathrm N}}=Q_{\overline{\mathrm p}} \hbar /(2 M_{\overline{\mathrm p}})$ is the antinuclear magneton, and \gl\ and \gs\ are the orbital and spin $g$-factors of the antiproton, respectively. Evidently, \gl\ is expected to be equal to 1, but this has never been measured for neither proton nor antiproton bound to an atom.

The electron magnetic moment is simply given by its spin part
\begin{equation}
\vec{\mu_{e}}=g^{e}_\ell \vec{L_{e}},
\end{equation}
and the interaction of the two magnetic moments leads to an unusual splitting: the two largest moments, the \pbar\ {\em angular} moment and the {\em electron} spin create the dominant splitting following $\vec{F} $= \Lp\ + \Se\ called {\em hyperfine} (HF) splitting, and the \pbar\ spin leads to a further splitting according to the total angular momentum $\vec{J}=\vec{F}$ + \Sp\ =  \Lp\ +  \Se\     + \Sp. The latter is called {\em superhyperfine} splitting (SHF), and the resulting quadruplet structure is shown in Fig.~\ref{EWfig:leveldiag} right.


The hyperfine structure of \pbhe\ has first been calculated by Bakalov and Korobov \cite{Bakalov:98}, who showed that the HF splitting is in the order of \nuHF = $10-15$ GHz for the metastable states, while the SHF splitting is two orders of magnitude smaller (\nuSHF\ = $100-300$ MHz). They furthermore showed that the difference of hyperfine splittings in laser transitions is very small for favoured transitions
with $\Delta v=0 $ (so-called because of the larger dipole transition moment), but exceeds the experimental resolution of $\sim 1$ GHz in unfavoured  $\Delta v=2 $ transitions. A scan of the $(n,L)=(37,35)\rightarrow(38,34)$ transition done in the last year of LEAR indeed revealed a doublet structure with a splitting of \DnuHF = $(1.70 \pm 0.5)$ GHz \cite{Widmann:97}, in accoradance with the theoretical value of 1.77 GHz \cite{Bakalov:98}.

In this experiment, the {\em difference} of the HF splittings of the two states $(37,35)$ and $(38,34)$ was measured to about 3\% precision. In order to directly observe HF transitions within one state $(n,L)$ and to determine the HF splitting to much higher precision, we devised a laser-microwave-laser resonance method which works as follows: The wavy lines in Fig.~\ref{EWfig:leveldiag} right represent allowed M1
transitions (flipping \Se\ but not \Sp ) which can be induced by
microwave radiation. All the HF levels are initially nearly equally
populated. In order to create a population asymmetry which is needed to
detect a microwave transition, a laser pulse at time $t_1$ stimulating a transition
from a metastable ($\tau \sim$ $\mu$s) state to a short-lived ($\tau
\lesssim $ 10 ns) state can be used. When the \pbar\ is excited to the
short-lived state, the \pbhe undergoes an Auger transition to a
\pbheion\ ion which is immediately destroyed via collisional
Stark-effect in the dense helium medium followed by annihilation of the
\pbar\ with a nucleon. An second on-resonance laser pulse at $t_2$ therefore superposes
a sharp spike onto the analog delayed annihilation spectrum (ADATS) whose area is
proportional to the population of the metastable state at the time of
the arrival of the laser pulse.



\begin{figure}
\centering
\includegraphics[width=6cm]{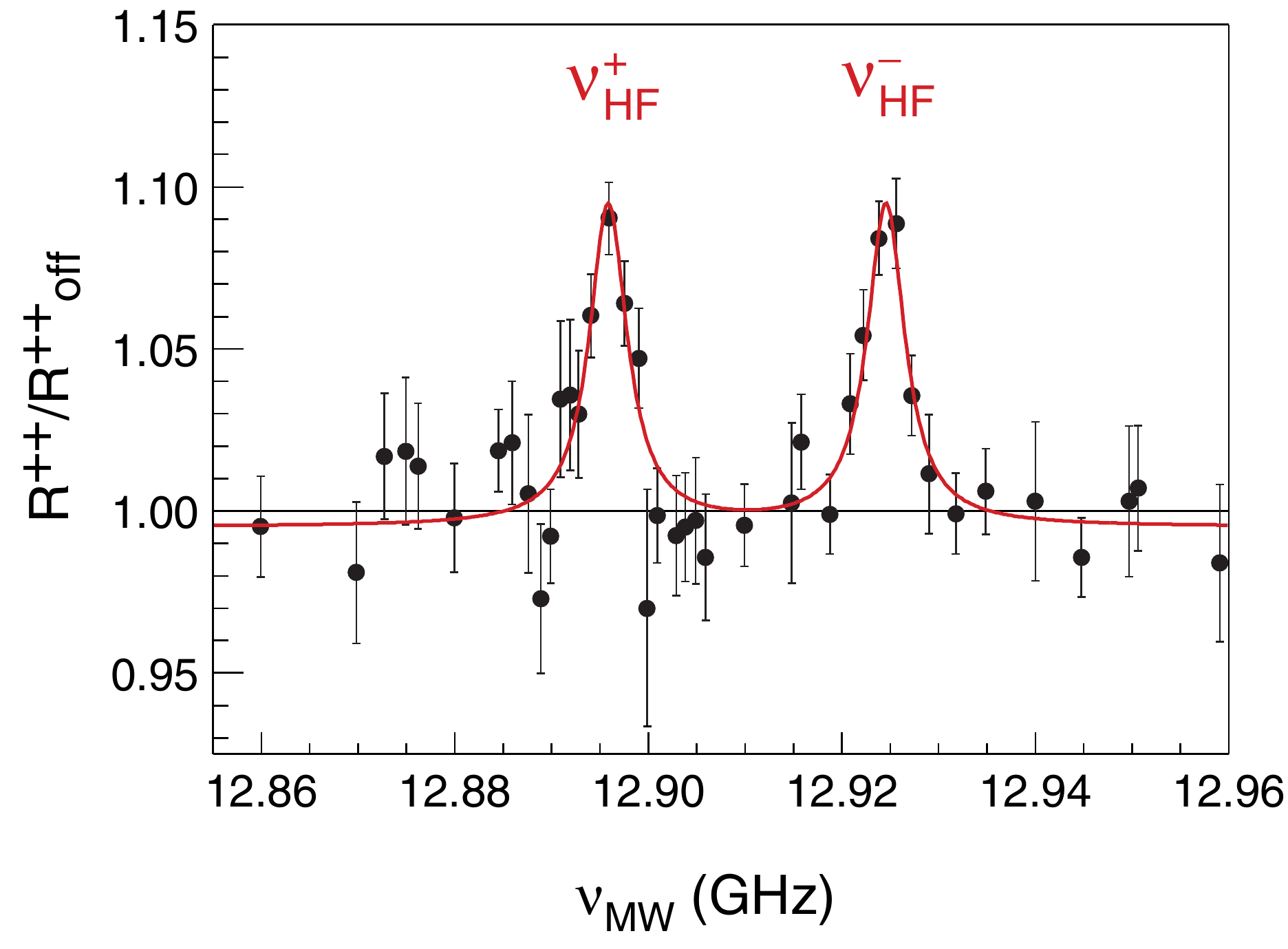}
\includegraphics[width=5.25cm]{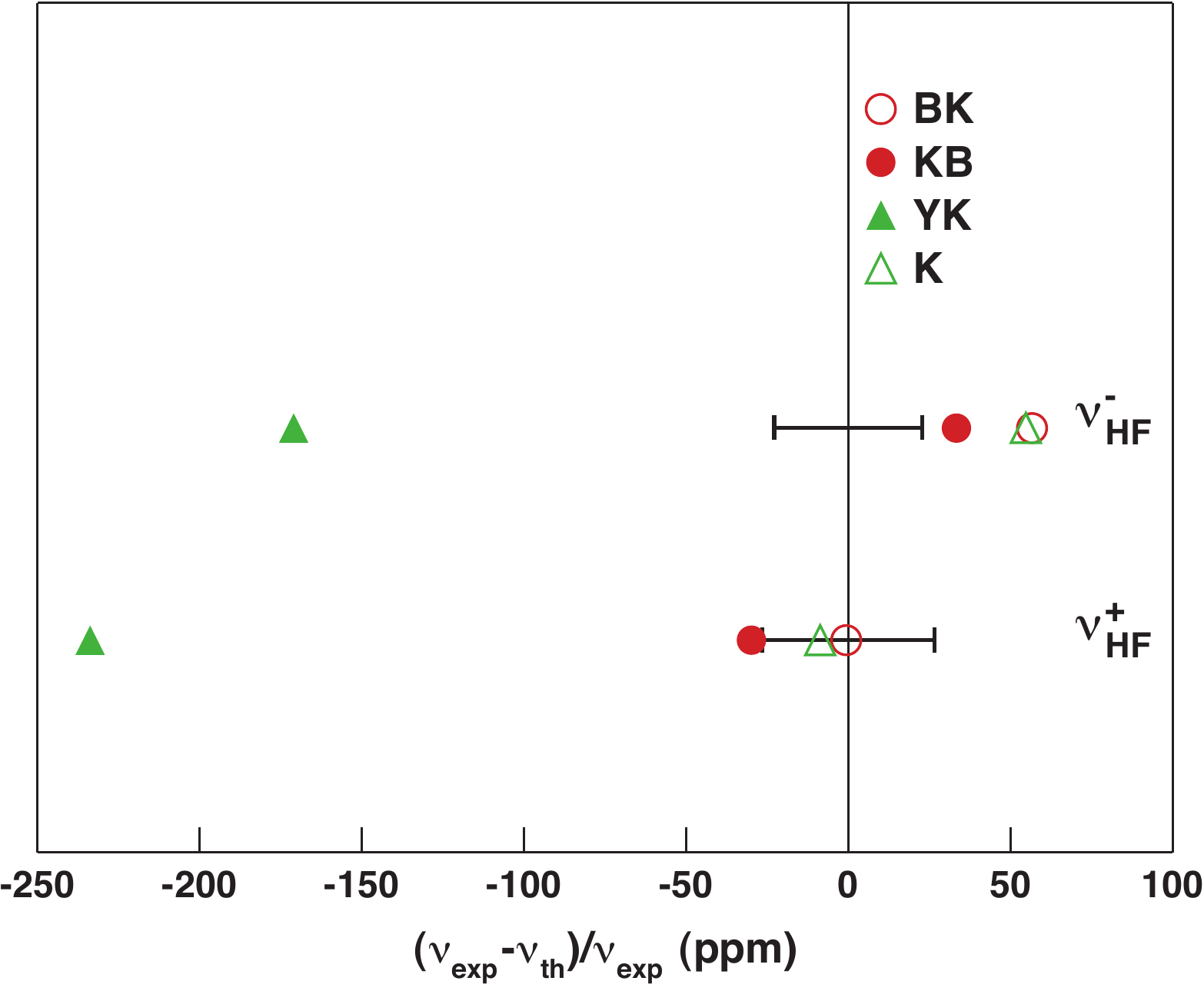}
\caption{Left: Experimental result of the laser-microwave-laser experiment \cite{Widmann:02}. right: Comparison of the experimental result with different calculations. Plottet is the relative deviation of theory and experiment in ppm. BK: \cite{Bakalov:98}, KB: \cite{Korobov:01}, YK: \cite{Yamanaka:01}, K: \cite{Kino:03APAC}.}
\label{EWfig:HFSresults}       
\end{figure}

The experiment has been performed at the AD and the result indeed showed the expected two resonance lines (cf. Fig.~\ref{EWfig:HFSresults} left, \cite{Widmann:02}). The two lines were measured with a relative accuracy of $\sim 3 \times 10^{-5}$ and agreed with most theoretical calculations at a level of $\sim 6 \times 10^{-5}$, which is comparable to the expected accuracy of the calculations originating from the omission of terms of relative order $\alpha^2 \approx 5 \times 10^{-5}$ (cf. Fig.~\ref{EWfig:HFSresults} right). The width of the resonance lines of $(5.3 \pm 0.7)$ MHz  is given by the Fourier limit caused by the time distance of $\Delta t = t_2 - t_1 = 160$ ns  between the two laser pulses, leading to an expected width of $\Delta \nu_{\mathrm MW} = 1/ \Delta t =6.25$ MHz.

The agreement between theory and experiment can be used to constrain the values of the magnetic moment of the antiproton. Since the observed transitions \nuHF $^\pm$ involve a spin flip of the electron, they are primarily sensitive  to the orbital magnetic moment of \pbar, {\em i.e. } the orbital $g$-factor \gl. 
The results imply $|$\gl $-1| < 6 \times 10^{-5}$. This result is unique since no corresponding value for the proton has ever been measured due to the absence of protonic atoms in our matter-dominated world.
 
The spin magnetic moment \mup\ (or equivalently \gs) is of greater importance for tests of CPT, since it is a direct property of the antiproton, and its value so far is known only to a precision of $\sim 0.3$\% \cite{PDG:06} from the measurement of X-rays of antiprotonic lead \cite{Kreissl:88}. As recently shown by Bakalov and Widmann \cite{Bakalov:07}, the sensitivity of the \nuHF $^\pm$ transitions on \mup\ is rather small and their measurement do not promise an improvement of the current PDG value. The difference \DnuHF\ = \nuHFm\ -- \nuHFp, however, is equal to the difference of SHF splittings \nuSHFp\ -- \nuSHFm\ and is therefore directly sensitive to \mup. The experimental error in \DnuHF\ is much larger than the one of  \nuHF $^\pm$, and using a sensitivity factor from \cite{Bakalov:07}, the current experimental precision corresponds to an uncertainty in \mup\ of $\sim 1.6$\%.

An improvement of the experimental precision is only possible if the line width can be reduced, {\em i.e.} the laser pulse distance $\Delta t$ can be prolonged. Using the new pulse-amplified cw-laser developed by ASACUSA in 2004, first tests were performed in 2006 showing that this is indeed possible. An improvement of the experimental precision of  an order of magnitude would be expected, which should allow to determine \mup\ to 0.1\% or better.

\section{Summary}

Antiprotonic helium continues to provide one of the most precise tests of CPT in the baryon sector. For charge and mass of the antiproton, an accuracy of 2 ppb has been reached and the use of Doppler-free two-photon spectroscopy promises further improvement. The hyperfine structure has been determined with high precision which will be increased in the near future to provide a value of the antiproton magnetic moment which will be more precise than the current PDG value.


\newcommand{\SortNoop}[1]{}

\end{document}